\documentclass[a4paper,12pt]{article}
\usepackage[T1]{fontenc} 
\usepackage[russian, english]{babel} 
\usepackage[dvips]{graphicx} 
\usepackage{epsfig}
\usepackage[dvips]{texdraw} \input{txdtools}
\usepackage{picinpar} 
\usepackage{amsmath}
\usepackage{amssymb}
\usepackage{indentfirst}
\usepackage{bm}

\newcommand{\n}{\mathbf{n}}

\DeclareMathOperator{\e}{\mathnormal{e}}
\DeclareMathOperator{\diag}{diag}

\begin{document}

\begin{center}
\textbf{Gravitational four-fermion interaction and dynamics of the early Universe}
\end{center}

\begin{center}
I.\,B. Khriplovich$^{a, b}$ \footnote{khriplovich@inp.nsk.su} and A.\,S. Rudenko$^{a, b}$ \footnote{a.s.rudenko@inp.nsk.su}
\end{center}
\begin{center}
$^a$ Budker Institute of Nuclear Physics, 630090, Novosibirsk, Russia \\
$^b$ Novosibirsk State University, 630090, Novosibirsk, Russia
\end{center}

\vspace{1mm}

\begin{abstract}

If torsion exists, it generates gravitational four-fermion interaction (GFFI), essential on the Planck scale. We analyze the influence of this interaction on the Friedmann-Lema\^{\i}tre-Robertson-Walker (FLRW) cosmology. Explicit analytical solution is derived for the problem where both the energy-momentum tensor generated by GFFI and the common ultrarelativistic energy-momentum tensor are included. We demonstrate that gravitational four-fermion interaction does not result in Big Bounce.
\end{abstract}

\vspace{5mm}

\textbf{1. Introduction}

\vspace{3mm}

According to the common belief, present expansion of the Universe is the result of Big Bang. One more quite popular idea is that this expansion had been preceded by the compression with subsequent Big Bounce. We analyze here the assumption that the Big Bounce is due to the gravitational four-fermion interaction.

The observation that, in the presence of (non-propagating) torsion, the interaction of fermions with gravity results in the four-fermion interaction of axial currents, goes back at least to \cite{ki, ro}.

The most general form of the gravitational four-fermion interaction (GFFI) is as follows: 
\begin{multline} \label{ff} 
S_{ff} = \frac{3 \pi}{2} \frac{\gamma^2}{\gamma^2 + 1} \, G \int d^4 x \, \sqrt{- g} \ \eta_{I J} \, \times \\
\times \left[ \left( 1 - \beta^2 + \frac{2 \beta}{\gamma} \right) A^I A^J - 2 \alpha \left( \beta - \frac{1}{\gamma} \right) V^I A^J - \alpha^2 V^I V^J \right];
\end{multline}
here and below $G$ is the Newton gravitational constant; $g$ is the determinant of the metric tensor; $A^I$ and $V^I$ are the total axial and vector neutral currents, respectively: 
\begin{equation} \label{A} 
A^I = \sum_a A_a^I = \sum_a \bar \psi_a \, \gamma^5 \, \gamma^I \, \psi_a \, , \qquad V^I = \sum_a V_a^I = \sum_a \bar \psi_a \, \gamma^I \, \psi_a \, ,
\end{equation}
the sums over $a$ in \eqref{A} extend over all elementary fermions with spin 1/2. $\alpha$, $\beta$ and $\gamma$ are numerical parameters of the problem. The values of $\alpha$ and $\beta$ are unknown. As to the so-called Barbero-Immirzi parameter $\gamma$, we assume the value $\gamma = 0.274$ \cite{khk}. In fact, the exact numerical values of these parameters are inessential for our problem.

The $AA$ contribution to expression \eqref{ff} corresponds (up to a factor) to the action derived long ago in \cite{ki, ro}. Then, this contribution was obtained in the limit $\beta \to 0$, $\gamma \to \infty$ in \cite{ke}. The present form of the $AA$ interaction, given in \eqref{ff}, was derived in \cite{pe, kib}. The $VV$ and $VA$ terms in \eqref{ff} were derived in \cite{fr} and \cite{kib, fr}, respectively.

Simple dimensional arguments demonstrate that interaction \eqref{ff}, being proportional to the Newton constant $G$ and to the particle number density squared, $n^2$, could get essential and comparable to the common interactions only at very high densities, i.\,e. on the Planck scale.

Quite extensive list of references on the papers, where the GFFI is discussed in connection with cosmology, can be found in \cite{kib, sh, be}.

\vspace{5mm}

\textbf{2. Energy-momentum tensor}

\vspace{3mm}

The energy-momentum tensor (EMT) $T_{\mu \nu}^{ff}$ generated by action \eqref{ff} is 
\begin{equation} \label{EMT} 
T_{\mu \nu}^{ff} = - \frac{3 \pi \gamma^2 G}{2 (\gamma^2 + 1)} \, g_{\mu \nu} \, \eta_{I J} \left[ \left( 1 - \beta^2 + \frac{2 \beta}{\gamma} \right) A^I A^J - 2 \alpha \left( \beta - \frac{1}{\gamma} \right) V^I A^J - \alpha^2 V^I V^J \right]. 
\end{equation}
The nonvanishing components of this expression, written in the locally inertial frame, are energy density $T_{00}^{ff} = \rho_{ff}$ and pressure $T_{11}^{ff} = T_{22}^{ff} = T_{33}^{ff} = p_{ff}$ (they are marked here and below by $ff$ to indicate their origin from the four-fermion interaction; for the correspondence between $\rho$, $p$ and EMT components see \cite{ll}, \S\,35).

Let us analyze the expressions for $\rho_{ff}$ and $p_{ff}$ in our case of the interaction of two ultrarelativistic fermions (labeled $a$ and $b$) in their locally inertial center-of-mass system. We follow here the line of reasoning of \cite{kh}.

The axial and vector currents of fermion $a$ are, respectively,
\begin{align}
A^I_a & = \frac{1}{4 E^2} \, \phi^\dagger_a \left\{ E \, \bm \sigma_a (\bm p' + \bm p), \; (E^2 - \bm p' \bm p) \, \bm \sigma_a + \bm p' (\bm \sigma_a \bm p) + \bm p \, (\bm \sigma_a \bm p') - i \, [\bm p' \times \bm p] \right\} \phi_a = \nonumber \\
& = \frac{1}{4} \, \phi^\dagger_a \, \bigl\{ \bm \sigma_a (\bm n' + \bm n), \; (1 - \bm n' \bm n) \, \bm \sigma_a + \bm n' (\bm \sigma_a \bm n) + \bm n \, (\bm \sigma_a \bm n') - i \, [\bm n' \times \bm n] \bigr\} \, \phi_a; \\
V^I_a & = \frac{1}{4 E^2} \, \phi^\dagger_a \left\{ E^2 + \bm p' \bm p + i \, \bm \sigma_a \, [\bm p' \times \bm p], \; E \, \bigl( \bm p' + \bm p - i \, \bm \sigma_a \times (\bm p' - \bm p) \bigr) \right\} \phi_a = \nonumber \\
& = \frac{1}{4} \, \phi^\dagger_a \, \bigl\{ 1 + \bm n' \bm n + i \, \bm \sigma_a \, [\bm n' \times \n], \; \bm n' + \bm n - i \, \bm \sigma_a \times (\bm n' - \bm n) \bigr\} \, \phi_a;
\end{align}
here $E$ is the energy of fermion $a$, $\phi_a$ is two-component spinor, $\bm n$ and $\bm n'$ are the unit vectors of its initial and final momenta $\bm p$ and $\bm p'$, respectively; under the discussed extreme conditions all fermion masses can be neglected. In the center-of-mass system, the axial and vector currents of fermion $b$ are obtained from these expressions by changing the signs: $\bm n \to - \bm n$, $\bm n' \to - \bm n'$. Then, after averaging over the directions of $\bm n$ and $\bm n'$, we arrive at the following semiclassical expressions for the nonvanishing components of the energy-momentum tensor, i.\,e. for the energy density $\rho_{ff}$ and pressure $p_{ff}$:\footnote{Note, that in Refs. \cite{kh, kr} the $AA$ contribution to $\rho_{ff}$ and $p_{ff}$ is 4 times smaller. AR thanks S.\,K. Maity for the query that led to a recalculation of $AA$ and $VV$ contributions.}
\begin{align}
\rho_{ff} = T_{00}^{ff} & = - \frac{\pi \gamma^2 G}{12 (\gamma^2 + 1)} \sum_\text{$a, b$} n_a n_b \left[ \left( 1 - \beta^2 + \frac{2 \beta}{\gamma} \right) \Bigl( 3 - 11 \langle \bm \sigma_a \bm \sigma_b \rangle \Bigr) - \alpha^2 \Bigl( 15 - 7 \langle \bm \sigma_a \bm \sigma_b \rangle \Bigr) \right] \nonumber \\
& = - \frac{\pi}{12} \, \frac{\gamma^2}{\gamma^2 + 1}\, G n^2 \left[ \left( 1 - \beta^2 + \frac{2 \beta}{\gamma} \right) \Bigl( 3 - 11 \, \zeta \Bigr) - \alpha^2 \Bigl( 15 - 7 \, \zeta \Bigr) \right];
\end{align}
\begin{align}
p_{ff} & = T_{11}^{ff} = T_{22}^{ff} = T_{33}^{ff} = \nonumber \\
& = \frac{\pi \gamma^2 G}{12 (\gamma^2 + 1)} \sum_\text{$a, b$} n_a n_b \left[ \left( 1 - \beta^2 + \frac{2 \beta}{\gamma} \right) \Bigl( 3 - 11 \langle \bm \sigma_a \bm \sigma_b \rangle \Bigr) - \alpha^2 \Bigl( 15 - 7 \langle \bm \sigma_a \bm \sigma_b \rangle \Bigr) \right] = \nonumber \\
& = \frac{\pi}{12} \, \frac{\gamma^2}{\gamma^2 + 1}\, G n^2 \left[ \left( 1 - \beta^2 + \frac{2 \beta}{\gamma} \right) \Bigl( 3 - 11 \, \zeta \Bigr) - \alpha^2 \Bigl( 15 - 7 \, \zeta \Bigr) \right];
\end{align}
here and below $n_a$ and $n_b$ are the number densities of the corresponding sorts of fermions and antifermions, $n = \sum_a n_a$ is the total density of fermions and antifermions, the summation $\sum_\text{$a, b$}$ extends over all sorts of fermions and antifermions; $\zeta = \,\langle\bm \sigma_a\bm \sigma_b\rangle$ is the average value of the product of corresponding $\bm \sigma$-matrices, presumably universal for any $a \neq b$. Since the number of sorts of fermions and antifermions is large, one can neglect here for numerical reasons the contributions of exchange and annihilation contributions, as well as the fact that if $\bm \sigma_a$ and $\bm \sigma_b$ refer to the same particle, $\langle\bm \sigma_a\bm \sigma_b\rangle = 3$. It is only natural that after the performed averaging over all momenta orientations, the $P$-odd contributions of $VA$ to $\rho_{ff}$ and $p_{ff}$ vanish.

Thus, the equation of state (EOS) is here
\begin{equation} \label{EOS}
\rho_{ff} = - p_{ff} = - \frac{\pi}{12} \, \frac{\gamma^2}{\gamma^2 + 1}\, G n^2 \left[ \left( 1 - \beta^2 + \frac{2 \beta}{\gamma} \right) \Bigl( 3 - 11 \, \zeta \Bigr) - \alpha^2 \Bigl( 15 - 7 \, \zeta \Bigr) \right]. 
\end{equation}
The four-fermion energy density \eqref{EOS} can be conveniently rewritten as
\begin{equation} \label{EOS1} 
\rho_{ff} = \varepsilon \, G n^2, \quad \text{with } \ \varepsilon = - \frac{\pi}{12} \, \frac{\gamma^2}{\gamma^2 + 1} \left[ \left( 1 - \beta^2 + \frac{2 \beta}{\gamma} \right) \Bigl( 3 - 11 \, \zeta \Bigr) - \alpha^2 \Bigl( 15 - 7 \, \zeta \Bigr) \right]. 
\end{equation}
Parameter $\zeta = \langle \bm \sigma_a \bm \sigma_b \rangle$ for $a \neq b$, just by its physical meaning, in principle can vary in the interval from 0 (which corresponds to the complete thermal incoherence or to the antiferromagnetic ordering) to 1 (which corresponds to the complete ferromagnetic ordering). Correspondingly, $\varepsilon$ varies from 
\begin{equation}
\varepsilon = - \frac{\pi}{4} \, \frac{\gamma^2}{\gamma^2 + 1} \left( 1 - \beta^2 + \frac{2 \beta}{\gamma} - 5 \alpha^2 \right) \quad \text{for } \ \zeta = 0 
\end{equation} 
to 
\begin{equation} 
\varepsilon = \frac{2 \pi}{3} \, \frac{\gamma^2}{\gamma^2 + 1} \left( 1 - \beta^2 + \frac{2 \beta}{\gamma} + \alpha^2 \right) \quad \text{for } \ \zeta = 1. 
\end{equation}
The absolute numerical value of the parameter $\varepsilon$ is inessential for the analysis below. Its sign, however, is crucial for the physical implications, and depends on $\alpha, \beta$ and $\zeta$. As to $\zeta$, most probably, at the discussed extreme conditions of high densities and high temperatures, this correlation function is negligibly small.

We go over now to the contributions of common matter to the energy density and pressure. For the extreme densities, where GFFI gets essential, this matter is certainly ultrarelativistic, and its contribution to the energy density can be written, for simple dimensional reasons, as 
\begin{equation} \label{EOSt} 
\rho = \nu \, n^{4/3},
\end{equation} 
where $\nu$ is a numerical factor. One power of $n^{1/3}$ is here an estimate for the energy per particle. Another factor $n$ in this expression is the total density of ultrarelativistic particles and antiparticles, fermions and bosons, contributing to \eqref{EOSt}. Since bosons also contribute to the total energy density, this factor should exceed the fermion density $n$ entering the above four-fermion expressions. This difference, however, is absorbed in \eqref{EOSt} by the factor $\nu$. As it was the case with $\rho_{ff}$, it is natural to assume that $\rho$ as well is independent of the spin correlations.

Let us consider now the EMT of the common ultrarelativistic matter in our problem. Since the problem is isotropic, the mixed components of the EMT should vanish:
\[
T_{0m} = T_{m0} = 0 \qquad (m = 1, 2, 3).
\]
Then, the space components of the EMT can be diagonalized, and due to the same isotropy, we arrive at
\[
T_{11} = T_{22} = T_{33} \, .
\]
At last, the trace of the EMT of this ultrarelativistic matter should vanish, $T^\mu_\mu = 0$. Thus, the discussed EMT can be written as 
\begin{equation} \label{com} 
T^\mu_\nu = \rho \, \diag \left( 1, - \frac{1}{3}, - \frac{1}{3}, - \frac{1}{3} \right) \quad \text{ or } \quad T_{\mu \nu} = \rho \, \diag \left( 1, \frac{1}{3}, \frac{1}{3}, \frac{1}{3} \right); 
\end{equation}
here and below $\rho$ is the energy density of the common ultrarelativistic matter, and its pressure is $p = \rho/3$.

With $\rho_{ff} \sim G n^2$, close to the Planck scale GFFI is quite comparable to $\rho \sim n^{4/3}$, so that on this scale both contributions should be included. Unfortunately, in our previous papers on the subject the contribution of the common ultrarelativistic matter was not taken into account.

\vspace{5mm}

\textbf{3. FLRW equations}

\vspace{3mm}

We will assume that, even on the scale close to Planck one, the Universe is homogeneous and isotropic, and thus can be described by the Friedmann-Lema\^{\i}tre-Robertson-Walker (FLRW) metric 
\begin{equation} \label{fr}
ds^2 = dt^2 - a(t)^2 \left[ dr^2 + f(r) \left( d\theta^2 + \sin^2 \theta\, d\phi^2 \right) \right]; 
\end{equation} 
here $f(r)$ depends on the topology of the Universe as a whole:
\[
f(r) = r^2, \quad \sin^2 r, \quad \sinh^2 r
\]
for the spatial flat, closed, and open Universe, respectively.

Now the total energy density and total pressure are, respectively,
\[ 
\rho_{tot} = \rho_{ff} + \rho, \qquad p_{tot} = - \rho_{ff} + \frac{1}{3} \, \rho \, .
\] 
The fact that $\rho_{ff}$ and $\rho$ enter the expression for the total pressure with opposite signs can be traced back to the difference between algebraic structures of the tensors $T^{ff}$ and $T$. The first of them is proportional in the mixed components to $\delta^\mu_\nu = \diag \left( 1, 1, 1, 1 \right)$, and the second one is proportional to $\diag \left( 1, - 1/3, - 1/3, - 1/3 \right)$.

Thus, the Einstein equations for the FLRW metric \eqref{fr} are 
\begin{align}
\left( \frac{\dot a}{a} \right)^2 + \frac{k}{a^2} = \frac{8 \pi G}{3} \, \rho_{tot} & = \frac{8 \pi G}{3} \left( \rho_{ff} + \rho \right), \label{a1} \\
\frac{\ddot a}{a} = - \frac{4 \pi G}{3} \left( \rho_{tot} + 3 p_{tot} \right) & = \frac{8 \pi G}{3} \left( \rho_{ff} - \rho \right); \label{a2}
\end{align} 
parameter $k$ in equation \eqref{a1} equals 0, 1, and $- 1$ for the spatial flat, closed, and open Universe, respectively. These equations result in the covariant conservation law for the total energy-momentum tensor: 
\begin{equation} \label{e}
\dot \rho_{tot} + 3 \, \frac{\dot a}{a} \left( \rho_{tot} + p_{tot} \right) = \dot \rho_{ff} + \dot \rho + 4 \, \frac{\dot a}{a} \, \rho = 0 \, . 
\end{equation}

Let us note here that in the absence of the four-fermion interaction, i.\,e. for $\rho_{ff} = 0$, this equation reduces to the well-known one for the common ultrarelativistic matter: $\dot \rho \, + 4 \, (\dot a/a) \, \rho = 0$\,.

On the other hand, without the common matter, i.\,e. for $\rho = 0$, equation \eqref{e} degenerates into $\dot \rho_{ff} = 0$. This is quite natural since energy-momentum tensor \eqref{EMT}, generated by the four-fermion interaction, can be conserved by itself only with $\rho_{ff} = \text{const}$ \cite{kr}.

In fact, observational data strongly favor the idea that our Universe is spatial flat, i.\,e. that $k = 0$. Then equation \eqref{a1} simplifies to 
\begin{equation} \label{a0}
\left( \frac{\dot a}{a} \right)^2 = \frac{8 \pi G}{3} \left( \rho_{ff} + \rho \right). 
\end{equation}

Obviously, if the gravitational four-fermion interaction exists, our equations \eqref{a1}--\eqref{a0} are as firmly established as the common FLRW equations in the absence of the GFFI.

\vspace{5mm}

\textbf{4. Solutions and conclusions}

\vspace{3mm}

Let us go over now to the solution of FLRW equations. With substitution
\begin{equation} \label{sub} 
a(t) = a_0 \e^{f(t)}, 
\end{equation} 
equations \eqref{a2} and \eqref{a0} transform to
\begin{align}
& \frac{8 \pi G}{3} \left( \rho_{ff} + \rho \right) = \dot f \, ^2, \label{dot} \\ 
& \frac{8 \pi G}{3} \, \rho = - \frac{1}{2} \, \ddot f \, . \label{ddot} 
\end{align}

Now, differentiating equation \eqref{dot} over $t$ and combining the result with \eqref{ddot}, we arrive at the following solution: 
\begin{equation} \label{ft} 
f = - \frac{3}{4 \nu} \, \varepsilon \, G \, n^{2/3} - \frac{1}{3} \ln n.
\end{equation} 
The numerical factor $\nu$ was introduced in \eqref{EOSt}. A comment on the ratio $\varepsilon \, G/\nu$ in this expression is pertinent. It can be easily demonstrated that in the absence of the four-fermion interaction, it is just relation 
\[
f = - \frac{1}{3} \ln n,
\]
which results in the law $a(t) = \sqrt t $. Then, it is only natural that the relative weight of the four-fermion interaction enters formula \eqref{ft} via the ratio $\varepsilon \, G/\nu$.

Thus we obtain 
\begin{equation} \label{res} 
a(t) = a_0 \e^{f(t)} \sim n^{- 1/3} \, \exp \left\{ - \frac{3}{4 \, \nu} \, \varepsilon \, G \, n^{2/3} \right\}. 
\end{equation}

Let us introduce the dimensionless ratio $\xi (t)$ of the four-fermion energy density $\rho_{ff}$ and the energy density $\rho$ of ultrarelativistic matter:
\begin{equation} \label{ksi} 
\xi (t) = \frac{\rho_{ff}}{\rho} = \frac{\varepsilon \, G}{\nu} \, n^{2/3}. 
\end{equation}
Then 
\begin{equation} \label{res1} 
a(t) \sim \frac{1}{\sqrt{\mathstrut \xi (t)}} \, \exp \left\{ - \frac{3}{4} \, \xi (t) \right\}.
\end{equation}
Combining equations \eqref{dot} and \eqref{ft}, we arrive at 
\begin{equation} \label{eqn}
\dot \xi = \mp \, \frac{4}{3} \, \sqrt{\frac{8 \pi G}{3}} \, \frac{\nu^{\,3/2}}{\varepsilon \, G} \; \xi^2 \, \frac{\sqrt{\mathstrut \xi + 1}}{\xi + 2/3} \, , 
\end{equation} 
which results in relations between $\xi$ and $t$: 
\begin{equation} \label{ksi0}
\ln \left( \frac{\sqrt{\mathstrut \xi (t)}}{1 + \sqrt{\mathstrut 1 + \xi (t)}} \right) - \frac{1}{2} \, \frac{\sqrt{\mathstrut 1 + \xi (t)}}{\xi (t)} = \mp \, \sqrt{\frac{8 \pi G}{3}} \, \frac{\nu^{\, 3/2}}{\varepsilon \, G} \, t + \text{const} \quad \text{for } \ \varepsilon > 0,
\end{equation}
and
\begin{equation} \label{ksi1}
- \ln \left( \frac{\sqrt{\mathstrut | \xi (t) |}}{1 + \sqrt{\mathstrut 1 - | \xi (t) |}} \right) - \frac{1}{2} \, \frac{\sqrt{\mathstrut 1 - | \xi (t) |}}{| \xi (t) |} = \mp \, \sqrt{\frac{8 \pi G}{3}} \, \frac{\nu^{\, 3/2}}{| \varepsilon | \, G} \, t + \text{const} \quad \text{for } \ \varepsilon < 0.
\end{equation}
The constants on right-hand sides of \eqref{ksi0} and \eqref{ksi1} are fixed by initial conditions. As to the signs in formulas \eqref{eqn}, \eqref{ksi0} and \eqref{ksi1}, $-$ and $+$ therein refer to expansion and compression, respectively.

The physical implications of formula \eqref{res} for positive and negative values of parameter $\varepsilon$ are quite different.

\begin{figure}[h]
\center
\begin{tabular}{c c c}
\includegraphics[scale=0.7]{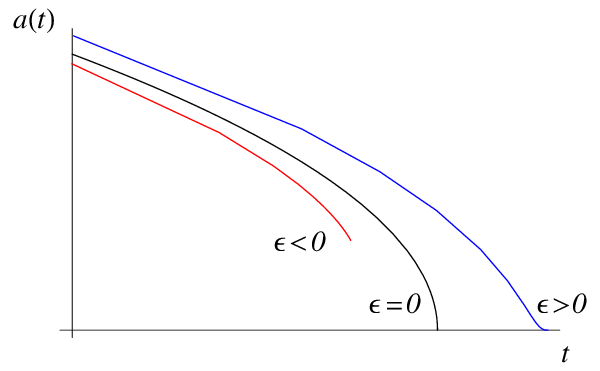} & &
\includegraphics[scale=0.7]{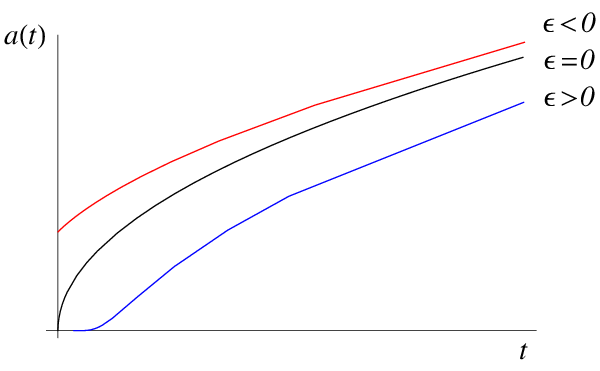} \\
a (compression) & & b (expansion)
\end{tabular}
\caption{\label{fig:1} (color online). Time-dependence of scale factor}
\end{figure}

For positive $\varepsilon$, both factors in \eqref{res}, $n^{- 1/3}$ and $\exp \{ - 3/(4 \, \nu) \, \varepsilon \, G \, n^{2/3} \}$, and of course their product $a(t)$, shrink to zero together with increasing density $n$. To analyze the compression, we rewrite equations \eqref{a2} and \eqref{a0} as follows: 
\begin{align} 
\dot a & = - \sqrt{\frac{8 \pi G}{3}} \, a \, \sqrt{\rho_{ff} + \rho}, \\
\ddot a & = \frac{8 \pi G}{3} \, a \, \left( \rho_{ff} - \rho \right).
\end{align} 
At the initial moment, when $\rho_{ff} \ll \rho$, both $\dot a$ and $\ddot a$ are negative, therefore the Universe shrinks with acceleration. Then at $\rho_{ff} = \rho$ acceleration $\ddot a$ changes sign, while $\dot a$ remains negative, therefore the compression of the Universe decelerates. According to relations \eqref{res} and \eqref{ksi0}, it takes finite time for $a$ to shrink to zero. Due to the exponential factor in \eqref{res}, $\dot a$ and $\ddot a$ also vanish at the same moment (the curve $\varepsilon > 0$ in Fig.~\ref{fig:1}a). Therefore, repulsive GFFI does not stop the collapse, but only reduces its rate. The asymptotic behavior of $a(t)$ is 
\begin{equation} 
a(t) \sim \left( t_1 - t \right) \exp \left\{ - \frac{9 \, \varepsilon^2 \, G}{128 \, \pi \, \nu^3} \, \frac{1}{\left( t_1 - t \right)^2} \right\};
\end{equation} 
here $t_1$ is the moment of the collapse for $\varepsilon > 0$.

For negative $\varepsilon$, the situation is different. Here the right-hand side of \eqref{res1}
\[
a(t) \sim \frac{1}{\sqrt{\mathstrut | \xi (t) |}} \, \exp \left\{ \frac{3}{4} \, | \xi (t) | \right\}
\]
reaches its minimum value at $| \xi_m | = 2/3$, i.\,e., $a(t)$ cannot decrease further. It follows from \eqref{a0}, however, that the compression rate $\dot a$ at this point does not vanish and remains finite (the curve $\varepsilon < 0$ in Fig.~\ref{fig:1}a). In a sense, the situation here resembles that in the standard cosmology with ultrarelativistic particles: therein $a(t) \sim \sqrt{\mathstrut t_0 - t} \to 0$ for $t \to t_0$ ($t_0$ is the moment of the collapse in this case), though at this point $\dot a$ does not vanish, but tends to infinity (the curve $\varepsilon = 0$ in Fig.~\ref{fig:1}a). In the standard cosmology one does not expect that this compression to the origin is followed by expansion. Therefore, in the present case, with $\varepsilon < 0$, it looks natural to assume as well that the compression will not change to expansion.

Thus, contrary to possible na\"{\i}ve expectations \cite{kh}, the gravitational four-fermion interaction does not result in Big Bounce.

We note in conclusion that it is difficult (if possible at all) to imagine a realistic possibility to detect any effect of the gravitational four-fermion interaction.

\subsection*{Acknowledgements}

We are grateful to V.V. Sokolov for a useful discussion.

The investigation was supported in part by the Foundation for Basic Research through Grant No. 11-02-00792-a, by the Ministry of Education and Science of the Russian Federation, and by the Grant of the Government of Russian Federation, No. 11.G34.31.0047.


\begin{thebibliography}{99}

\bibitem{ki} 
T.\,W.\,B. Kibble,
\textit{Lorentz invariance and the gravitational field}, 
J. Math. Phys. \textbf{2} (1961) 212.

\bibitem{ro} 
V.\,I. Rodichev, 
\textit{Twisted space and nonlinear field equations},
J. Exp. Theor. Phys. \textbf{40} (1961) 1469.

\bibitem{khk} 
I.\,B. Khriplovich and R.\,V. Korkin,
\textit{How is the maximum entropy of a quantized surface related to its area?}, 
J. Exp. Theor. Phys. \textbf{95} (2002) 1 [arXiv:gr-qc/0112074].

\bibitem{ke} 
G.\,D. Kerlick, 
\textit{Cosmology and particle pair production via gravitational spin-spin interaction in the Einstein-Cartan-Sciama-Kibble theory of gravity}, 
Phys. Rev. \textbf{D 12} (1975) 3004.

\bibitem{pe} 
A. Perez and C. Rovelli, 
\textit{Physical effects of the Immirzi parameter},
Phys. Rev. \textbf{D 73} (2006) 044013 [arXiv:gr-qc/0505081].

\bibitem{kib} 
J. Magueijo, T.\,G. Zlosnik, and T.\,W.\,B. Kibble, 
\textit{Cosmology with a spin}, 
Phys. Rev. \textbf{D 87} (2013) 063504 [arXiv:1212.0585 [astro-ph.CO]].

\bibitem{fr} 
L. Freidel, D. Minic, and T. Takeuchi,
\textit{Quantum gravity, torsion, parity violation and all that}, 
Phys. Rev. \textbf{D 72} (2005) 104002 [arXiv:hep-th/0507253].

\bibitem{sh} 
I.\,L. Shapiro, 
\textit{Physical aspects of the space-time torsion}, 
Phys. Rep. \textbf{357} (2002) 113 [arXiv:hep-th/0103093].

\bibitem{be} 
G. de Berredo-Peixoto, L. Freidel, I.\,L. Shapiro, and C.\,A. de Souza, 
\textit{Dirac fields, torsion and Barbero-Immirzi parameter in Cosmology}, 
JCAP \textbf{06} (2012) 017 [arXiv:1201.5423 [gr-qc]].

\bibitem{ll} 
L.\,D. Landau and E.\,M. Lifshitz, 
\textit{The Classical Theory of Fields}, 
Butterworth-Heinemann (1975).

\bibitem{kh} 
I.\,B. Khriplovich, 
\textit{Big bounce and inflation from gravitational four-fermion interaction}
in V.\,G. Gurzadyan, A. Klumper and A.\,G. Sedrakyan eds., 
\textit{Low Dimensional Physics and Gauge Principles: Matinyan Festschrift}, 
World Scientific, (2012), pg. 183 [arXiv:1203.5875 [gr-qc]].

\bibitem{kr} 
I.\,B. Khriplovich and A.\,S. Rudenko, 
\textit{Cosmology constrains gravitational four-fermion interaction},
JCAP \textbf{11} (2012) 040 [arXiv:1210.7306 [astro-ph.CO]].

\end{thebibliography}
\end{document}